\documentclass[useAMS,usenatbib]{mn2e}
\usepackage{graphicx}

\newcommand{\ms}{$\,$M$_\mathrm{\odot}$}
\newcommand{\ls}{$\,$L$_\mathrm{\odot}$}
\newcommand{\be}{\begin{equation}}
\newcommand{\ee}{\end{equation}}
\newcommand{\stars}{{\sc stars}}
\newcommand{\el}[2]{\ensuremath{^{#1}\mathrm{#2}}}
\newcommand{\eps}[1]{\ensuremath{\epsilon(\mathrm{#1})}}

\newcommand{\pa}{\ensuremath{\mathrm{(p,\alpha)}}}
\newcommand{\ag}{\ensuremath{\mathrm{(\alpha,\gamma)}}}
\newcommand{\an}{\ensuremath{\mathrm{(\alpha,n)}}}
\newcommand{\np}{\ensuremath{\mathrm{(n,p)}}}

\newcommand{\gr}{\ensuremath{\bigtriangledown_\mathrm{rad}}}
\newcommand{\gad}{\ensuremath{\bigtriangledown_\mathrm{ad}}}
\title[Mass loss and yield uncertainty in AGB stars]{Mass loss and yield uncertainty in low-mass asymptotic giant branch stars}
\author[R.~J. Stancliffe, C.~S. Jeffery]{Richard J. Stancliffe$^1$\thanks{E-mail:
rs@ast.cam.ac.uk} and C. Simon Jeffery$^2$\\
$^1$Institute of Astronomy, The Observatories, Madingley Road, Cambridge CB3 0HA, U.K. \\
$^2$Armagh Observatory, College Hill, Armagh BT61 9DG, Northern Ireland}
\begin{document}
\bibliographystyle{mn2e}

\date{Accepted 0000 December 00. Received 0000 December 00; in original form 0000 October 00}

\pagerange{\pageref{firstpage}--\pageref{lastpage}} \pubyear{0000}

\maketitle

\label{firstpage}

\begin{abstract}
We investigate the uncertainty in surface abundances and yields of asymptotic giant branch (AGB) stars. We apply three different mass loss laws to a 1.5\ms\ star of metallicity $Z=0.008$ at the beginning of the thermally pulsing asymptotic giant branch (TP-AGB) phase. Efficient third dredge-up is found even at very low envelope mass, contrary to previous simulations with other evolution codes. We find that the yield of carbon is uncertain by about 15\% and for most other light elements the yield is uncertain at the level of 20-80\%. For iron group elements the uncertainty varies from around 30\% for the more abundant species to over a factor of two for the less abundant radioactive species, like \el{60}{Fe}. The post-AGB surface abundances for this mass and metallicity are much more uncertain due to the dilution of dredged-up material in differing envelope masses in the later stages of the models. Our results are compared to known planetary nebula (PN) and post-AGB abundances. We find that the models are mostly consistent with observations but we are unable to reproduce observations of some of the isotopes.
\end{abstract}


\section{Introduction}
Stars of between about 1 and 8\ms\ experience a phase of unstable double shell burning known as thermal pulses whilst on the asymptotic giant branch (AGB). During this phase, the star experiences significant mass loss, which strips away the envelope to leave behind a degenerate core of carbon and oxygen. The mechanism for this mass loss is unknown and this lack of knowledge is a major source of uncertainty in calculations of thermally pulsing asymptotic giant branch (TP-AGB) evolution. Mass loss allows the nucleosynthetic products of AGB evolution to be returned to the interstellar medium and, because low-mass stars are formed far more abundantly than those of higher mass \citep{1993MNRAS.262..545K}, allows AGB stars to contribute significantly to galactic chemical evolution. Differences in mass-loss history may affect the evolution and nucleosynthesis of a star and hence lead to different yields.

Mass loss also has important effects on the evolution of AGB stars. In the lowest-mass AGB stars it is possible for mass loss to strip the star of its envelope before third dredge-up (TDUP) has a chance to occur \citep{2002PASA...19..515K}. If TDUP does occur its efficiency can still be affected by mass loss. In intermediate-mass AGB stars the occurrence of mass loss can inhibit the occurrence of hot-bottom burning in the later stages of evolution, while still allowing TDUP to occur. This allows the formation of bright carbon stars \citep{1998A&A...332L..17F}.

Of the early calculations of AGB evolution, few used any mass loss. Those that did \citep[e.g.][]{1979A&A....79..108S} employed the Reimers' formula \citep{1975psae.book..229R}
\be
\dot{M} = - 4 \times 10^{-13}\eta {LR\over M} \mathrm{M_\odot}\mathrm{yr}^{-1}
\label{eq:reimers}
\ee
where $L,R$ and $M$ are the star's luminosity, radius and mass in solar units. The value of the free parameter $\eta$ is usually taken as 0.4 for red giants. However, it was recognized by \citet{1981pprg.work..431R} that the mean mass-loss rate required to produce a typical planetary nebula was around $3\times10^{-5}$\ms$\mathrm{yr}^{-1}$. He coined the term superwind as this value was significantly greater than those given by the Reimers formula. Many calculations still use the Reimers formula, generally with $0.4<\eta<3$, although $\eta=10$ has been used in extreme cases \citep{1997ApJ...478..332S,2003PASA...20..389S}. 

In an attempt to produce a more physically consistent picture, \citet{1993ApJ...413..641V} produced a mass-loss relation based on observations of OH/IR stars and other pulsating, dust-enshrouded AGB stars. They linked the mass-loss rate, $\dot{M}$ in \ms$\mathrm{yr}^{-1}$, of an object to its Mira pulsational period, $P$ (in days), via the relations:
\be
\log \dot{M} = -11.4 + 0.0123P
\ee
and
\be
\dot{M} = {L\over cv_\mathrm{exp}},
\ee
where $v_\mathrm{exp} = -13.5 + 0.056P$ is the expansion velocity of the wind far from the star in km\,s$^{-1}$ and $c$ is the speed of light. The mass loss is taken to be the smaller of the two values calculated with the above expressions. This law has the advantage of providing for a superwind phase which can remove a star's envelope over the course of a few pulses. This mass loss law has been used in recent computations by \citet{2002PASA...19..515K}. 

One other mass-loss relation has also been used in extant calculations, that of \citet{1995A&A...297..727B}. It is grounded in theory, rather than empirically determined, based as it is on simulations of shock-driven winds in the atmospheres of Mira-like stars \citep{1988ApJ...329..299B}. It was developed to reflect the strong increase of mass loss during the AGB and to be applicable to stellar evolution calculations \citep{1995A&A...297..727B}. It is similar to the Reimers' prescription but with a steeper dependence on luminosity and a shallower dependence on mass. The formula is expressed as:
\be
\dot{M} = 4.83\times10^{-9}\left(M\over\mathrm{M_\odot}\right)^{-2.1}\left(L\over\mathrm{L_\odot}\right)^{2.7}\dot{M}_\mathrm{Reimers}.
\ee
Because of its use of the Reimers' formula, this law also suffers from the problem of having a free parameter that must be determined. Recent calculations using Bl\"{o}cker's relation include those of \citet{2004ApJ...613L..73H} and \citet{2004MmSAI..75..654V}.

In this work, we address the uncertainty in the evolution and nucleosynthesis of a low-mass TP-AGB star due to our lack of understanding of the mass loss mechanism.

\section{The stellar evolution code \stars}
We use the stellar evolution code \stars\ which is a variant of the evolution code originally developed by \citet{1971MNRAS.151..351E}, and updated by many authors \citep[e.g.][]{1995MNRAS.274..964P}. The version used here includes the AGB-specific modifications of \citet*{2004MNRAS.352..984S}, together with the updated opacity tables of \citet{2004MNRAS.348..201E}. In order to study light element nucleosynthesis in detail we employ the nucleosynthesis subroutines of \citet{2005MNRAS.360..375S} which cover isotopes from deuterium to sulphur and important iron group elements -- a total of 40 isotopes.

A 1.5\ms\ model of metallicity $Z=0.008$ was evolved from the pre-main sequence using 199 mesh points. Initial abundances are solar-scaled and taken from the work of \citet{1989GeCoA..53..197A}. On the first giant branch and on the early asymptotic giant branch (E-AGB) \citet{1975psae.book..229R} mass loss was used with $\eta=0.4$. At the end of core helium burning, the model was remeshed to 499 mesh points to facilitate the transition to the high resolution (999 mesh points), AGB-specific mesh spacing function of \citet{2004MNRAS.352..984S}. Convective overshooting was not employed during any stage in the evolution.

On the TP-AGB, we use $\eta = 1$ when applying Reimers' mass loss. When applying Bl\"ocker's law, we use $\eta = 0.02$, following \citet*{2000A&A...363..605V} who chose this value based on calibration to the lithium luminosity function in the Large Magellanic Cloud.

\section{Results} 

\begin{figure}
\includegraphics[width=7.5cm]{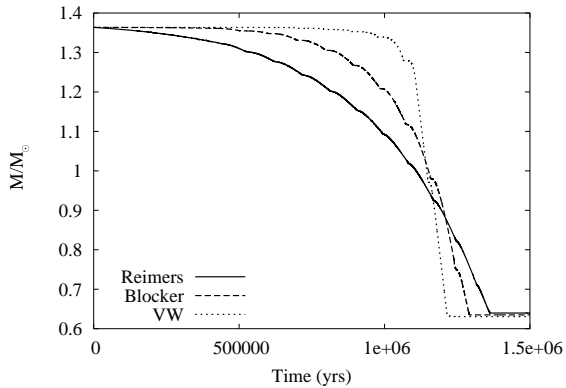}
\caption{Plot of the total stellar mass as a function of time from the beginning of the evolution run. The model using Reimers' mass loss is the solid line. The dashed line is the model with Bl\"ocker's mass loss and the dotted line is the model with Vassiliadis and Wood (VW) mass loss.}
\label{fig:massloss}
\end{figure}

The three mass-loss laws lead to different mass loss histories for each of the models. The evolution of stellar mass with time is shown in Figure~\ref{fig:massloss}. The model with Reimers' mass loss has appreciable mass loss from the beginning of the run. The rate of mass loss slowly increases with time before the envelope is completely removed. The model with Bl\"ocker's mass loss only begins to show appreciable mass loss after about $5\times10^5$\,yr. The mass loss then increases more rapidly as the model evolves. The model with Vassiliadis and Wood (hereafter VW) mass loss only shows appreciable mass loss once the superwind phase begins at around $10^6$\,yr.

\begin{figure}
\includegraphics[width=7.5cm]{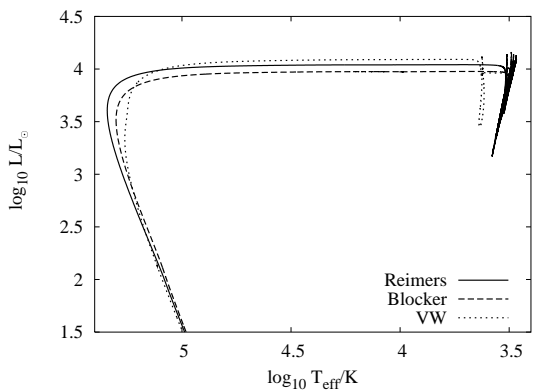}
\caption{Hertzsprung-Russell diagram for each of the three models. The model with VW mass loss experiences an AGB final thermal pulse.}
\label{fig:HR}
\end{figure}

When the model using VW mass loss reaches an envelope mass of around $3\times10^{-3}$\ms, the star experiences its last thermal pulse. Because the envelope mass is very low, the pulse occurs when the star is slightly further to the blue in the HR diagram (see Figure~\ref{fig:HR}). This is an AGB final thermal pulse \citep[AFTP, see][]{2001Ap&SS.275....1B}. 

Details of the evolution of the three models are presented in Tables~\ref{tab:reimerslambda}, \ref{tab:blockerlambda} and \ref{tab:VWlambda}. The Reimers' model undergoes the most thermal pulses but all models have just six thermal pulses followed by episodes of third dredge-up. Dredge-up begins slightly later in the the Reimers model, when the core mass has reached $0.602$\ms\ compared with $0.595$\ms\ and $0.594$\ms\ in the Bl\"ocker and VW models respectively. In terms of the total amount of intershell material dredged-up, all the models are comparable. The Reimers' model dredges up $0.0179$\ms\ of material while for the Bl\"ocker and VW models, $0.0171$\ms\ and $0.0176$\ms\ of intershell material are dredged up respectively. Because the Reimers model typically has a lower envelope mass for a given thermal pulse, when dredge-up occurs the material that is carried to the envelope does not get as diluted as it does in the other two models so a higher C/O ratio tends to be obtained (excluding the AFTP as this is a slightly different phenomenon). 

\begin{table*}
\begin{center}
\begin{tabular}{cccccccccc}
TP & $M_*$ & $M_\mathrm{env}$ & M$_\mathrm{H}$ & $\tau_\mathrm{ip}$ & $\log$ (L$^{\mathrm{max}}_\mathrm{He}$/\ls) & $\Delta M_\mathrm{H}$ & $\Delta M_\mathrm{DUP}$ & $\lambda$ & C/O \\ 
   & (\ms) & (\ms) & (\ms) & ($10^4$\,yr) & & (\ms) & (\ms) & & \\
\hline
1 & 1.36357 & 0.79285 & 0.57072 &  ...  & 4.63556 & 0.01112 &  ...  &  ...  & 0.324 \\ 
2 &  1.30354 & 0.71765 & 0.58183 & 7.94 & 6.42507 & 0.00405 &  ...  &  ...  & 0.324 \\ 
3 &  1.27895 & 0.68858 & 0.58587 & 9.57 & 6.48659 & 0.00449 &  ...  &  ...  & 0.324 \\ 
4 &  1.24563 & 0.64962 & 0.59035 & 9.67 & 6.89601 & 0.00566 &  ...  &  ...  & 0.324 \\ 
5 &  1.20506 & 0.60288 & 0.59599 & 9.59 & 7.04725 & 0.00619 &  ...  &  ...  & 0.324 \\ 
6 & 1.15642 & 0.54764 & 0.60203 & 9.30 & 7.20978 & 0.00675 & 0.00134 & 0.199 & 0.553 \\ 
7 & 1.09658 & 0.48161 & 0.60744 & 9.02 & 7.40975 & 0.00753 & 0.00252 & 0.335 & 1.086 \\ 
8 & 1.02219 & 0.40142 & 0.61245 & 8.80 & 7.60488 & 0.00832 & 0.00373 & 0.448 & 1.993 \\ 
9 & 0.92994 & 0.30389 & 0.61704 & 8.52 & 7.79146 & 0.00902 & 0.00303 & 0.336 & 2.860 \\ 
10 & 0.82767 & 0.19597 & 0.62303 & 7.84 & 7.81919 & 0.00867 & 0.00308 & 0.355 & 4.148 \\ 
11 & 0.71431 & 0.07700 & 0.62862 & 7.21 & 7.88218 & 0.00872 & 0.00422 & 0.484 & 8.072 \\ 
\hline
\end{tabular}
\end{center}
\caption{Details of the model computed using Reimers' mass loss with $\eta = 1$. The data are TP -- the thermal pulse number, $M_*$ -- the total stellar mass, $M_\mathrm{env}$ -- the envelope mass, $M_\mathrm{H}$ -- the hydrogen free core mass, $\tau_{\mathrm{ip}}$ -- the interpulse period, L$^\mathrm{max}_\mathrm{He}$ -- the peak luminosity from helium burning, $\Delta M_\mathrm{H}$ -- the hydrogen free core mass growth during the interpulse, $\Delta M_\mathrm{DUP}$ -- the mass of material dredged up, $\lambda = \Delta M_\mathrm{DUP}/\Delta M_\mathrm{H}$ -- the dredge-up efficiency and C/O -- the surface carbon-to-oxygen ratio by number.}
\label{tab:reimerslambda}
\end{table*}

\begin{table*}
\begin{center}
\begin{tabular}{cccccccccc}
TP & $M_*$ & $M_\mathrm{env}$ & M$_\mathrm{H}$ & $\tau_\mathrm{ip}$ & $\log$ (L$^{\mathrm{max}}_\mathrm{He}$/\ls) & $\Delta M_\mathrm{H}$ & $\Delta M_\mathrm{DUP}$ & $\lambda$ & C/O \\ 
   & (\ms) & (\ms) & (\ms) & ($10^4$\,yr) & & (\ms) & (\ms) & & \\
\hline
1 & 1.36357 & 0.79285 & 0.57072 &  ...  & 6.48915 & 0.01468 &  ...  &  ...  & 0.324 \\ 
2 & 1.34865 & 0.75917 & 0.58538 & 9.68 & 6.31532 & 0.00410 &  ...  &  ...  & 0.324 \\ 
3 & 1.33271 & 0.73750 & 0.58947 & 9.82 & 6.91354 & 0.00575 &  ...  &  ...  & 0.324 \\ 
4 & 1.30801 & 0.70657 & 0.59518 & 9.88 & 7.05616 & 0.00626 & 0.00054 & 0.086 & 0.362 \\ 
5 & 1.27055 & 0.66274 & 0.60090 & 9.48 & 7.23129 & 0.00691 & 0.00162 & 0.234 & 0.589 \\ 
6 & 1.21258 & 0.59870 & 0.60619 & 9.21 & 7.43439 & 0.00769 & 0.00274 & 0.356 & 1.048 \\ 
7 & 1.12323 & 0.50370 & 0.61114 & 8.93 & 7.61591 & 0.00839 & 0.00367 & 0.437 & 1.754 \\ 
8 & 0.98724 & 0.36246 & 0.61586 & 8.59 & 7.75249 & 0.00892 & 0.00451 & 0.506 & 2.873 \\ 
9 & 0.77352 & 0.14381 & 0.62027 & 8.26 & 7.91912 & 0.00945 & 0.00402 & 0.425 & 5.104 \\ 
\hline
\end{tabular}
\end{center}
\caption{Details of the model computed using Bl\"ocker's mass loss with $\eta = 0.02$. The columns are the same as in Table~\ref{tab:reimerslambda}.}
\label{tab:blockerlambda}
\end{table*}

\begin{table*}
\begin{center}
\begin{tabular}{cccccccccc}
TP & $M_*$ & $M_\mathrm{env}$ & M$_\mathrm{H}$ & $\tau_\mathrm{ip}$ & $\log$ (L$^{\mathrm{max}}_\mathrm{He}$/\ls) & $\Delta M_\mathrm{H}$ & $\Delta M_\mathrm{DUP}$ & $\lambda$ & C/O \\ 
   & (\ms) & (\ms) & (\ms) & ($10^4$\,yr) & & (\ms) & (\ms) & & \\
\hline
1 & 1.36357 & 0.79285 & 0.57072 &  ...  & 6.45469 & 0.01467 &  ...  &  ...  & 0.324 \\ 
2 & 1.36337 & 0.77396 & 0.58537 & 9.45 & 6.30073 & 0.00403 &  ...  &  ...  & 0.324 \\ 
3 & 1.36293 & 0.76796 & 0.58939 & 9.59 & 6.86964 & 0.00557 &  ...  &  ...  & 0.324 \\ 
4 & 1.36047 & 0.75949 & 0.59494 & 9.64 & 7.00581 & 0.00603 & 0.00055 & 0.091 & 0.359 \\ 
5 & 1.35569 & 0.74849 & 0.60042 & 9.33 & 7.18123 & 0.00678 & 0.00175 & 0.258 & 0.587 \\ 
6 & 1.34349 & 0.73042 & 0.60545 & 9.10 & 7.39886 & 0.00762 & 0.00314 & 0.412 & 1.047 \\ 
7 & 1.28632 & 0.66795 & 0.60993 & 8.91 & 7.60311 & 0.00844 & 0.00393 & 0.466 & 1.652 \\ 
8 & 0.98573 & 0.36262 & 0.61444 & 8.43 & 7.68659 & 0.00867 & 0.00432 & 0.498 & 2.798 \\ 
9 & 0.63154 & 0.00362 & 0.61879 & 8.22 & 7.83213 & 0.00914 & 0.00403 & 0.441 &28.01 \\ 
\hline
\end{tabular}
\end{center}
\caption{Details of the model computed using Vassiliadis and Wood mass loss. The columns are the same as in Table~\ref{tab:reimerslambda}. Note that the last pulse is an AGB final thermal pulse.}
\label{tab:VWlambda}
\end{table*}

\subsection{Dredge-up at low envelope mass}
One important feature of these models is the occurrence of third dredge-up even at very low envelope mass. Previous studies have suggested that TDUP should cease below envelope masses of around $0.5$\ms\ \citep{2003PASA...20..389S}. Calculations with the \stars\ code have given deeper dredge-up in low-mass stars compared with other simulations \citep{2005MNRAS.356L...1S}. The reason for this is unclear. Recent work by \citet{2006MNRAS.370.1817S} has shown that this is not due to the use of a simultaneous method of solution of the equations of stellar structure and evolution.

The physical cause for the occurrence of TDUP has received little attention in recent years. It is briefly touched upon by \citet{1976ApJ...208..165I} who states: \begin{quote}The entropy content of the convective envelope above the hydrogen-helium interface is immense relative to the entropy content of the helium-rich region below the interface. Hence, the absorption of comparable amounts of energy by each region corresponds to a much, much lower increase in entropy in the convective region above the interface than in the radiative region below the interface. The only way in which the continuity of the entropy parameter $\bar{S}$ can be maintained across the interface is for the interface to move inward in mass.\end{quote} This description is rather brief.

We offer the following description of why TDUP occurs. Energy from helium burning causes expansion throughout the star. This expands the hydrogen burning shell and extinguishes it. The outermost envelope initially contracts because it is no longer being supported by H-burning -- this explains the immediate dip in both surface luminosity and radius seen after a thermal pulse. Eventually, energy from He-burning filters through to the these regions and expansion resumes.

This expansion has the following effects. The expansion causes a drop in temperature and hence the opacity in the envelope increases. This causes a sharp rise in the radiative temperature gradient, \gr, pushing it above the relatively constant adiabatic gradient, \gad. This makes the region convective and leads to the occurrence of TDUP. The increase in entropy described by Iben is a consequence of the expansion, not a cause of third dredge-up.

Should we expect TDUP to continue even at low envelope masses? In the picture outlined above, the answer would seem to be `yes'. If sufficient energy from helium burning is dumped into the envelope, then TDUP should occur.

\begin{figure}
\includegraphics[width=7.5cm]{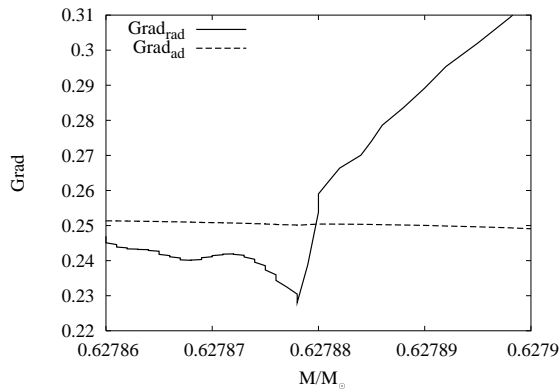}
\caption{The adiabatic and radiative temperature gradients as a function of mass at the base of the convective envelope. The adiabatic gradient, $\gad$, is the solid line. The dashed line represents the radiative gradient, $\gr$. This profile is taken from close to the beginning of TDUP in the last thermal pulse in the VW model. The plot is not smooth due to the number of decimal places that the output file uses to print the mass co-ordinate.}
\label{fig:gradprofile}
\end{figure}

Why do the models presented here give such efficient dredge-up at low envelope mass? Most importantly, the radiative temperature gradient does not show a discontinuity as it approaches the adiabatic gradient from the convective side of the border, as can be seen in Figure~\ref{fig:gradprofile}. This is an important thing to achieve in calculations of third dredge-up. A discontinuity in the radiative gradient would correspond to an unstable Schwarzchild boundary: mixing of material across this boundary would result in the movement of the location of this boundary \citep[see the discussion in][for example]{1999A&A...344..617M}. It is thus necessary to ensure that a stable boundary is obtained, as is done here, to ensure that dredge-up is accurately calculated.

It may be suggested that the depth of TDUP depends on the choice of the mixing coefficient at the edge of a convective zone. The \stars\ code employs an arithmetic mean to compute the mixing between two mesh points \citep[see the discussion in][]{2004MNRAS.352..984S}. It can be argued that this choice amounts to a form of convective overshooting as we are mixing material from beyond the formal (Schwarzchild) boundary for convective stability. However, we are mixing only one zone beyond the Schwarzchild boundary and this is much smaller that the scale of convective overshooting employed in a code like that of \citet{2003PASA...20..389S}, as the mass resolution around the base of the convective is of the order of  $10^{-6}$\ms. We also point out that \citet{2004MmSAI..75..676C} did not find efficient dredge-up at low envelope mass despite the inclusion of convective overshooting. The reason why we find deep dredge-up at low envelope mass without the use of convective overshooting while others do not (even though they include overshooting) is not understood.

\subsection{Nucleosynthesis}
Before discussing the nucleosynthesis of these models it is necessary to point out a deficiency. Because there is no convective overshooting (or other extra mixing process such as rotationally-driven mixing or internal gravity waves) used in the models, no carbon-13 pocket is formed. It is the reaction \el{13}{C}\an\el{16}{O} that is thought to be responsible for the $s$-process in low-mass AGB stars. This should be borne in mind when reactions involving neutron captures are discussed below.

\begin{figure*}
\includegraphics[angle=270,width=15cm]{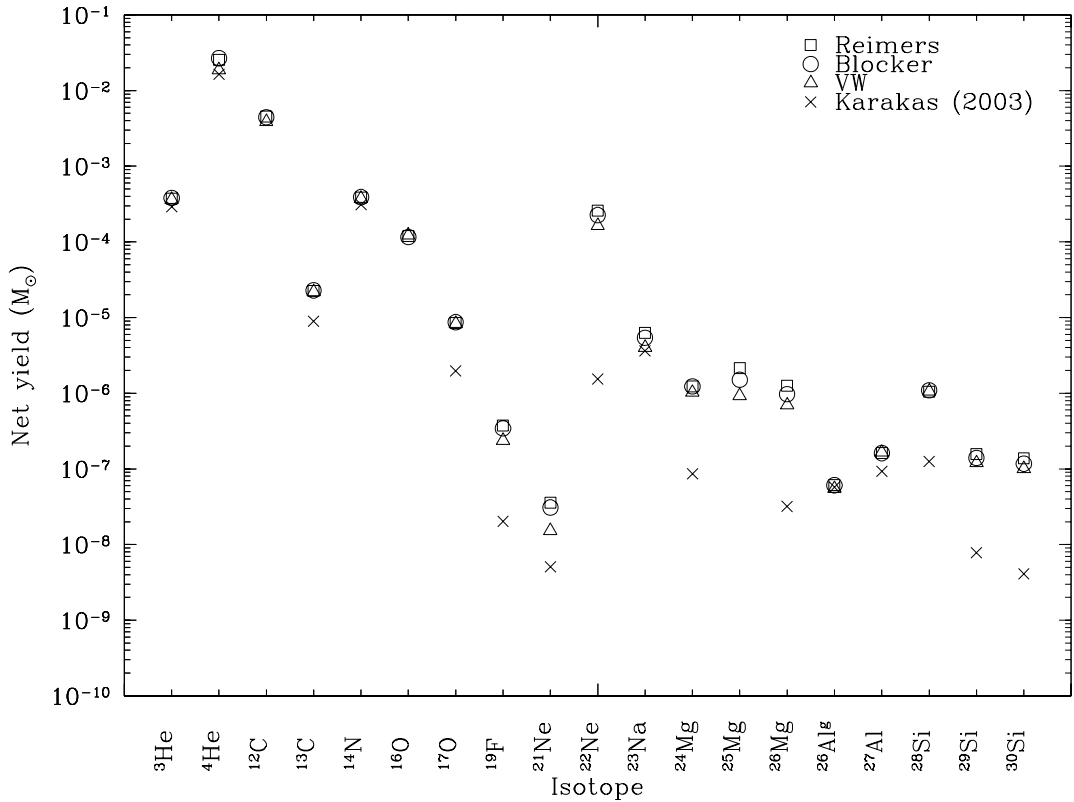}
\caption{Net yield in solar masses of selected isotopes from our models. Squares are yields from the model with Reimers' mass loss, circles are from the model with Bl\"ocker's mass loss and triangles are from the model with VW mass loss. For comparison, we have also included the yields of \citet{Karakas2003}, which are displayed with crosses. Only those isotopes with a positive net yield are displayed - hence the absence of yields for the isotopes \el{12}{C}, \el{16}{O} and \el{25}{Mg} from the Karakas model: her net yields for these isotopes are all negative.}
\label{fig:yield}
\end{figure*}

\begin{figure*}
\includegraphics[angle=270,width=15cm]{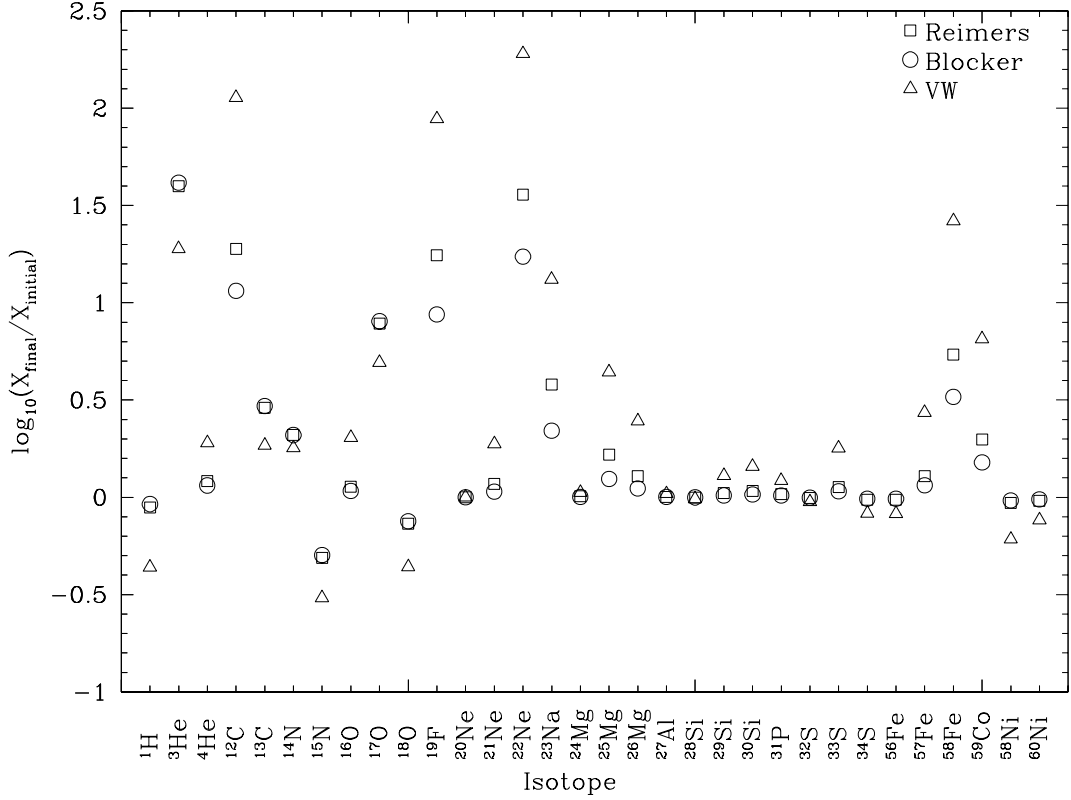}
\caption{Plot of the logarithm of the ratio of the final surface abundance to the initial surface abundance for selected isotopes in the nucleosynthesis network. Squares are from the model with Reimers' mass loss, circles are from the model with Bl\"ocker's mass loss and triangles are from the model with VW mass loss.}
\label{fig:abundance}
\end{figure*}

Light element yields and final surface abundances for the three models are presented in Tables~\ref{tab:lightyield} and \ref{tab:lightsurface}\footnote{Yields and abundances for all the isotopes in the nucleosynthesis network can be found in the Appendix.}. For ease of comparison, they are also displayed in Figures~\ref{fig:yield} and \ref{fig:abundance}. The signatures of nucleosynthesis along the whole TP-AGB are displayed most clearly in the model with Reimers' mass loss. This is because it typically has a lower envelope mass for each thermal pulse and so the dredged-up material suffers less dilution in the envelope. This explains why the surface abundances of some isotopes are significantly higher in this model compared with the model with Bl\"ocker mass loss. Because there is almost no envelope remaining when the VW model experiences its AFTP, its final envelope composition is dominated by the nucleosynthesis signatures of this last pulse cycle. 

Figure~\ref{fig:yield} also plots the yields of a 1.5\ms\ star of metallicity $Z=0.008$, as calculated by \citet{Karakas2003}, along side our yields. The Karakas model was calculated using the Mount Stromlo Stellar Structure Program \citep[see][and references therein]{Karakas2003}. It should be noted that the Karakas model uses the initial \el{12}{C}, \el{14}{N} and \el{16}{O} abundances of \citet{1992ApJ...384..508R}, rather than solar-scaled abundances as we have used. For isotopes like \el{14}{N}, whose abundance is mostly determined by evolution prior to the TP-AGB, we obtain very similar yields to those of the Karakas model. However, there are significant differences (of several orders of magnitude) between the yields of the Karakas model and the yields of our three models for those isotopes that are produced in the intershell of an AGB star. Most notable are the yields for \el{12}{C}, \el{16}{O} and \el{25}{Mg}. For these isotopes we obtain a positive net yield (see equation \ref{eq:netyield} below for a definition), whereas her model has a negative net yield. We attribute these differences to the fact that all of our models show quite efficient third dredge-up whereas hers does not.

When the models first reach $\log_{10} T_\mathrm{eff}=5.0$, the surface carbon abundance (by mass) of the model with Reimers' mass loss is $2.636\times10^{-2}$ compared to $1.602\times10^{-2}$ in the model with Bl\"ocker's mass loss and $1.579\times10^{-1}$ in the model with VW mass loss. However, because all three models undergo comparable amounts of dredge-up and lose roughly the same total mass, their carbon yields (i.e. material returned to the interstellar medium) are comparable, with a spread of just $6\times10^{-4}$\ms. This amounts to an uncertainty of about 15\% associated with the mass-loss law used. All yields described herein are net yields, i.e. 
\be
p = \int_0^\tau \dot{M}(t)[X_\mathrm{s}(t) - X_\mathrm{s}(0)]dt,
\label{eq:netyield}
\ee
where $\dot{M}(t)$ is the mass loss rate at time $t$, $X_\mathrm{s}(t)$ is the mass fraction of species s at time $t$ and $X_\mathrm{s}(0)$ is the initial mass fraction of species s in the material from which the star formed.

\begin{table*}
\begin{center}
\begin{tabular}{lccccccccc}
Mass loss & \el{12}{C} & \el{16}{O} & \el{19}{F} & \el{21}{Ne} & \el{22}{Ne} & \el{23}{Na} & \el{25}{Mg} & \el{26}{Mg} & \el{26}{Al}$^g$ \\
\hline
Reimers & 4.525(-3) & 1.197(-4) & 3.730(-7) & 3.571(-8) & 2.603(-4) & 6.301(-6) & 2.138(-6) & 1.255(-6) & 6.104(-8) \\
Bl\"ocker & 4.449(-3) & 1.155(-4) & 3.428(-7) & 3.091(-8) & 2.266(-4) & 5.419(-6) & 1.506(-6) & 9.753(-7) & 6.081(-8) \\
VW & 3.884(-3) & 1.234(-4) & 2.367(-7) &1.522(-8) & 1.646(-4) & 4.016(-6) & 9.254(-7) & 6.970(-7) & 5.498(-8) \\
\hline
Difference & 6.41(-4) & 7.90(-6) & 1.363(-7) & 2.05(-8) & 9.57(-5) & 2.29(-6) & 1.22(-6) & 5.58(-7) & 6.06(-9) \\
(\%) & 15.0 & 6.6 & 42.9 & 75.1 & 44.1 & 43.5 & 79.6 & 57.2 & 10.3 \\
\hline
\end{tabular}
\end{center}
\caption{Net yields (in solar masses) of selected light isotopes. All yields are expressed in the form $n(m) = n\times10^m$. The percentage difference is calculated using the mean of the three yields as a reference point.}
\label{tab:lightyield}
\end{table*}

\begin{table*}
\begin{center}
\begin{tabular}{lccccccccc}
Mass loss & \el{12}{C} & \el{16}{O} & \el{19}{F} & \el{21}{Ne} & \el{22}{Ne} & \el{23}{Na} & \el{25}{Mg} & \el{26}{Mg} & \el{26}{Al}$^g$ \\
\hline
Reimers &  2.636(-2) & 4.393(-3) & 2.859(-6) & 1.947(-6) & 1.884(-3) & 5.112(-5) & 4.514(-5) & 4.017(-5) & 2.031(-7) \\
Bl\"ocker &  1.602(-2) & 4.179(-3) & 1.418(-6) & 1.777(-6) & 9.074(-4) & 2.955(-5) & 3.384(-5) & 3.471(-5) & 1.314(-7) \\
VW & 1.579(-1) & 7.867(-3) & 1.436(-5) & 3.128(-6) & 9.985(-3) & 1.772(-4) & 1.200(-4) & 7.703(-5) & 9.671(-7) \\
\hline
Difference & 1.42(-1) & 3.69(-3) & 1.29(-5) & 1.35(-6) & 9.08(-3) & 1.47(-4)  & 8.61(-5) & 4.23(-5) & 8.36(-7) \\
(\%) & 212 & 67.3 & 208 & 59.1 & 719 & 171 & 130 & 83.6 & 193 \\
\hline
\end{tabular}
\end{center}
\caption{Surface abundances by mass fraction for selected light isotopes when the models reach $\log_{10} T_\mathrm{eff}=5.0$ for the first time. All abundances are expressed in the form $n(m) = n\times10^m$. The percentage difference is calculated using the mean of the three abundances as a reference point.}
\label{tab:lightsurface}
\end{table*}

The post-AGB \el{16}{O} abundance shows an uncertainty of around 67\%. Oxygen is produced in the intershell during helium burning and brought to the surface during third dredge-up. It is only a minor component of the intershell (at around 2\% by mass) and is little affected by reactions in the H-burning shell.  Hence there is little variation in the surface abundance (see Table~\ref{tab:lightsurface}), except in the VW model when the last pulse makes a significant contribution to the oxygen content of the envelope. The \el{16}{O} yield is well determined, with an uncertainty of 6.6\%. This is because the models all dredge-up a similar quantity of He-burning ashes. 

The pathway to \el{19}{F} production in AGB stars is a complicated one. It is produced from \el{14}{N} via  \citep{Goriely89}
\be
\el{14}{N}\ag\el{18}{F}(\beta^+)\el{18}{O}\pa\el{15}{N}\ag\el{19}{F}. \nonumber
\ee
In order to get p-capture on to \el{18}{O} in the H-depleted intershell, the \el{13}{C}\an\el{16}{O} reaction must be active in order to produce neutrons so that the reaction \el{14}{N}\np\el{14}{C} can occur. Uncertainties in fluorine production in AGB stars due to uncertainties in the reaction rates involved were studied by \citet{2004ApJ...615..934L}. They found an uncertainty of about 50\% in the yields from their low mass models. This is comparable to the uncertainty found here (43\%) due to the uncertainty in mass-loss rate. The post-AGB surface abundance of \el{19}{F} is found to be highly uncertain (by a factor of 3), with the model with VW mass loss having a much higher post-AGB surface abundance. The model with Reimers' mass loss also has a higher surface abundance than the Bl\"ocker's mass loss model. This is because in the VW model sequence, the dredged-up material suffers less dilution in the low mass of envelope remaining at the time of the last thermal pulse. The sequence using Bl\"ocker's mass loss has a more massive envelope at its last thermal pulse, so the dredged-up material suffers more dilution.

%This is likely to be due to this model experiencing more, hotter pulses and hence experiencing a greater neutron exposure from the marginal activation of the \el{22}{Ne}\an\el{25}{Mg} neutron source. The small envelope during these last pulses also helps to enhance the post-AGB surface abundance.

Of the Ne-Na cycle elements, \el{22}{Ne} is produced by $\alpha$-captures on to \el{14}{N} giving \el{18}{O} which subsequently captures another $\alpha$ particle to produce \el{22}{Ne}. This takes place in the intershell with \el{14}{N} being present in the ashes of H-burning. Once dredged-up, the \el{22}{Ne} experiences proton-captures in the H-burning shell and this affects the abundances of the other Ne-Na cycle elements (namely \el{20}{Ne}, \el{21}{Ne}, \el{23}{Na} as well as the two unstable isotopes \el{21}{Na} and \el{22}{Na}). The model with VW mass loss clearly displays the production of \el{22}{Ne} from the intershell and the Ne-Na cycle elements from the H-burning shell. There is even a trace amount of the radioactive isotope \el{22}{Na} in this model, with a surface abundance of $4\times10^{-10}$ when the star reaches a temperature of 100,000\,K.

The yields of \el{22}{Ne} and \el{23}{Na} have similar uncertainties (about 40\%). This is because the former comes from the intershell and all the models dredge-up similar amounts. \el{23}{Na} is formed from proton captures on to \el{22}{Ne} and hence closely follows its behaviour. \el{21}{Ne} is more sensitive to conditions in the burning shell and hence we see a greater uncertainty in its yield. In contrast, we see very large uncertainties in the abundances of \el{22}{Ne} and \el{23}{Na} (by factors of a few), compared to an uncertainty of 59\% in the \el{21}{Ne} abundance. This is because \el{21}{Ne} is a daughter of the abundant nucleus \el{20}{Ne} which does not experience much nucleosynthesis in these stars.  

For the heavier elements, the situation is more complicated. As the star proceeds along the TP-AGB its pulses get stronger, with the temperature at the base of the intershell convection zone (ICZ) getting higher. This opens up new nucleosynthesis pathways. Of particular interest is the reaction \el{22}{Ne}\an\el{25}{Mg}, which becomes active at temperatures of around $3\times10^{8}$\,K and provides a source of neutrons. We see evidence for this process in the yield and surface abundance of \el{25}{Mg} (and to a lesser extent, \el{26}{Mg}). The model with Reimers' mass loss has more, hotter pulses and hence a greater amount of \el{25}{Mg} is produced. The uncertainty in the yield due to the choice of mass-loss rate is around 80\%. This is over twice as great as the uncertainty associated with the \el{22}{Ne}$+\alpha$ reaction rates as found by \citet{2006ApJ...643..471K} who studied the production of heavy magnesium isotopes in intermediate-mass AGB stars.

Yields and post-AGB surface abundances for the iron group elements are shown in Tables~\ref{tab:ironyield} and \ref{tab:ironsurface}. These elements are only affected by neutron captures as their reaction rates with charged particles are extremely low at the temperatures found in AGB stars. The strongest depletion of \el{56}{Fe} is seen in the VW model as the post-AGB abundances of this model are dominated by the last thermal pulse. This is hot enough to allow substantial activation of the \el{22}{Ne}\an\el{25}{Mg} reaction. The \el{56}{Fe} abundance is very well determined because this species is very abundant and only a small portion of these nuclei suffer neutron captures. The yields are insignificant compared to the abundance of \el{56}{Fe} because this element does not suffer extensive nucleosynthesis on the TP-AGB.

\begin{table*}
\begin{center}
\begin{tabular}{lcccccc}
Mass loss & \el{30}{Si} & \el{56}{Fe} & \el{58}{Fe} & \el{60}{Fe} & \el{59}{Co} & \el{58}{Ni} \\
\hline
Reimers & 1.385(-7) & -1.604(-7) & 1.052(-6) & 6.607(-10) & 2.149(-7) & -1.303(-7) \\
Bl\"ocker & 1.169(-7) & -1.759(-7) & 9.963(-7) & 3.590(-10) & 2.113(-7) & -1.221(-7) \\
VW & 1.005(-7) & 3.672(-7) & 7.762(-7) & 1.025(-10) & 1.607(-7) & -8.098(-8) \\
\hline
Difference & 3.80(-8) & 5.276(-7) & 2.76(-7) & 5.58(-10) & 5.42(-8) & 4.93(-8) \\
(\%) & 32.0 & 44.2 & 29.3 & 149 & 27.7 & 14.8 \\
\hline
\end{tabular}
\end{center}
\caption{Net yield of selected iron group isotoptes in solar masses. For the per cent difference for \el{56}{Fe} the mean of the absolute values has been used as it is more representative.}
\label{tab:ironyield}
\end{table*}

\begin{table*}
\begin{center}
\begin{tabular}{lcccccc}
Mass loss & \el{30}{Si} & \el{56}{Fe} & \el{58}{Fe} & \el{60}{Fe} & \el{59}{Co} & \el{58}{Ni} \\
\hline
Reimers & 1.020(-5) & 4.563(-4) & 8.058(-6) & 6.439(-9) & 2.688(-6) & 1.858(-5) \\
Bl\"ocker & 9.785(-6) & 4.625(-4) & 4.883(-6) & 1.929(-9) & 2.047(-6) &  1.916(-5) \\
VW & 1.363(-5) & 3.882(-4) & 3.919(-5) & 2.647(-8) & 8.813(-6) &  1.211(-5) \\
\hline
Difference & 3.85(-6) & 7.43(-5) & 3.18(-6) & 2.45(-8) & 6.77(-6) & 7.05(-6) \\
(\%) & 34.3 & 17.1 & 197 & 211 & 150 & 42.4 \\
\hline
\end{tabular}
\end{center}
\caption{Post-AGB surface abundances by mass fraction of selected iron group isotopes.}
\label{tab:ironsurface}
\end{table*}

This picture is borne out by the rest of the iron group elements. We see evidence for greater processing of these elements by n-capture in the yield from the Reimers' model. This is because this model has the largest number of hot (T\ $>3\times10^8$K) pulses. There are large uncertainties in the post-AGB abundances of the n-capture progeny of \el{56}{Fe} because the initial abundances of these elements are low and hence small differences in neutron exposure lead to large differences in post-AGB abundance. The VW model shows the most processing of iron group elements in its post-AGB abundances because it is dominated by its last, hot TP.

The post-AGB surface abundances are more uncertain than the yields. This is because the post-AGB surface abundances depend on what nucleosynthesis is occurring and when it takes place relative to the loss of mass. If the envelope mass is lower when dredge-up occurs, the dredged-up material does not get diluted as much and hence the surface abundances can change more rapidly. The yields are less sensitive. The models lose comparable amounts of envelope in total so should display similar yields -- provided the same amount of material is dredged-up to the envelope. 

\section{Discussion}
%Having discussed the theoretical uncertainties we now move on to how the models compare to observations. There are two classes of object that can yield potential constraints on AGB evolution and nucleosynthesis. These are planetary nebula (PN) and post-AGB objects.
Having discussed the theoretical uncertainties we now move on to how the models compare to observations. While we could compare our models to observations of AGB star envelopes, we could not be sure how far such a star has gone through its TP-AGB evolution. We therefore restrict our comparison to planetary nebulae (PNe) and post-AGB objects, which we believe have completed their TP-AGB evolution.

\subsection{Planetary Nebulae}
Some planetary nebulae may be the result of AGB evolution, with the material lost during a  superwind phase being ionised by the hot CO core once it is exposed. Hence PN should display the signatures of AGB nucleosynthesis. However, we must be careful that the PN we compare come from objects of similar mass. For this reason, we do not compare these models to nitrogen-rich PN which presumably come from more massive stars that experience hot bottom-burning. \citet{2004A&A...426..779M} present abundance determinations for planetary nebula in the LMC. They find \eps{He}\,=\,10.96, \eps{N}\ $=7.46$, \eps{O}\ $=8.35$ and  \eps{S}\ $=6.81$, where \eps{X}\ $=\log (X/H) + 12$ with X and H being abundance by number. The errors on these measurements are of the order of 0.2-0.3 dex.

We compare the post-AGB abundances of our models to these values. The PN that could have formed from our models should have come from the final thermal pulses and hence have a composition not too dissimilar to the abundances of the central object. We therefore have \eps{He}\,$\simeq11.05$, \eps{N}\,$\simeq8.00$, \eps{O}\,$\simeq8.59$ and \eps{S}\,$=6.88$ for the Reimers' and Bl\"ocker models. Note that the variation between these two models is substantially less than the error in the measurements, being of the order of 0.01 dex. The agreement with the observations is reasonable but by no means perfect with the nitrogen abundance being somewhat off. This may be because we have assumed solar-scaled abundances which may not apply to the LMC. \citet{1992ApJ...384..508R} suggest that the nitrogen abundance of the LMC may be a factor of 4 lower than the solar-scaled abundance. The VW model, which has a significantly lower hydrogen abundance than the other two models, gives \eps{He}\ $=11.60$, \eps{N}\ $=8.25$, \eps{O}\ $=9.19$ and \eps{S}\ $=7.17$. Whether these abundances would be representative of the PN abundance is questionable as there is very little envelope left (after the AFTP) to be thrown off from this model.

Fluorine has also been recently detected in planetary nebula. \citet{2005ApJ...631L..61Z} presented measurements of \el{19}{F} abundances in PN which suggest that in a PN with C/O\,$\simeq2$, [\el{19}{F}/\el{16}{O}]\,$\simeq1$. They conclude that their data are consistent with the observations of \citet{1992A&A...261..164J} for giant star envelopes.
Our models are not consistent these observations on two counts. Firstly, our C/O ratios are much higher (by about 2.5 in the model with Bl\"ocker's mass loss and by about 4 in the model with Reimers' mass loss). In addition, the model [\el{19}{F}/\el{16}{O}] values are too low. The VW model gives [\el{19}{F}/\el{16}{O}]\,$= 1.64$, which is in agreement with the measurements. However, the C/O ratio for this model is an order of magnitude too great to match the observations.

\subsection{Post-AGB objects}
We now turn our attention to observations of post-AGB objects. Here we must stress that the theoretician's definition of a post-AGB object does not match that of an observer. To a theoretician, a post-AGB object is an object that was recently on the AGB and is now evolving from the AGB tip to the white dwarf cooling track. An observer's definition of a post-AGB object is a low-mass star that lies above the horizontal-branch and to the left of the giant branch. In the following, we compare theoretical and observed abundances for stars which we believe to be post-AGB stars in the theoretical sense.

\citet{2006A&A...450..701S} provide data on C, N, O and S abundances for 125 post-AGB objects. We ignore their data on R Coronae Borealis (RCrB) stars and extreme helium stars as these are unlikely to be post-AGB objects in the theoretical sense and hence we do not expect our models to fit them. 

Observation of sulphur should be dominated by \el{32}{S} and this undergoes little nucleosynthesis on the TP-AGB, being subject to limited neutron captures only. Our models suggest a value of \eps{S}$ = 6.9-7.2$. At this \eps{S}, the data suggest a maximum C/O ratio of about 2.4, whereas our models range from 5.1 to 28.0. 

For objects of 0.62\ms\ (which is approximately the mass of the post-AGB objects formed in the models), \citet{2006A&A...450..701S} find that the total abundance of CNO elements over the sulphur abundance (relative to solar) should lie between 0 and 0.8. Our models give [(C+N+O)/S]$=0.77$ which is consistent with the data. This measure should be dominated by the carbon abundance in a low-mass object because hot bottom burning, which converts carbon to nitrogen, is not active. The models are consistent with the oxygen abundance of the observations, though there is a large spread observed in the latter. Observations suggest \eps{O} should lie in the range 8.2-9.2. The models all fall within this range. The carbon abundances of the Reimers' and Bl\"ocker models are also consistent with the data, with \eps{C} in the range 9.23-9.53. This range is comparable to the data of \citet{2006A&A...450..701S} who do find objects in this range. The VW model is somewhat more C-rich.

We underline that we have only looked at one mass of AGB star, whereas the data should contain post-AGB objects that have come from AGB stars of a range of masses. It is this spread in masses (and other parameters, such as metallicity) that should give the observed spread in abundances. We are also assuming that we would expect to see post-AGB objects from this mass and metallicity of AGB star in the data set of \citet{2006A&A...450..701S}.

\subsection{H-deficient post-AGB objects}
The last thermal pulse of the VW model is qualitatively different from those of the other two models. Because of its low envelope mass, the hydrogen abundance of this model falls rapidly during TDUP. While this model does not become truly H-deficient, it is informative to compare it with the H-deficient post-AGB objects of the PG~1159 class. \citet{1998A&A...334..618D} provide data on the abundance of both pulsating and non-pulsating PG~1159 objects. Our model is too N-deficient to belong to the pulsating class as these objects have N/He ratios of around 0.01. However, it has a comparable C/He ratio to the non-pulsators, though it is probably too oxygen deficient. The non-pulsating PG~1159 objects in Table~3 of \citet{1998A&A...334..618D} have C/He ratios in the range 0.05-0.3 and O/He ratios in the range 0.005-0.1.  

\subsection{Additional processes}
Whilst the models are consistent with the main features of the observations, there are still some discrepancies which cannot be explained. The models all produce C/O ratios that are higher than both the observed PN and post-AGB object C/O ratios. There are three possible solutions to this problem:
\begin{itemize}
\item{TDUP is not as efficient (particularly in the later stages of the TP-AGB) as we find,}
\item{mass loss is stronger in the later stages so that the envelope is removed before the C/O ratio becomes too high, or}
\item{some extra mixing process \citep[see e.g.][]{1995ApJ...442L..21B}, which allows material in the envelope to mix to deeper in the star, may be active so that partial CN-cycling may take place.}
\end{itemize}
The latter of these options, which would result in a lowering of the C/O ratio, is particularly attractive as it would help to alleviate the other major problem associated with the models, namely the failure to match the observations of [\el{19}{F}/\el{16}{O}] versus C/O \citep[see the discussion in][for example]{2004ApJ...615..934L}. 

The failure of the models to match the \el{19}{F} observations may also be due to the fact that the models do not include a \el{13}{C} pocket. This deprives the models of an important source of neutrons and hence may lead to an underproduction of fluorine.

\section{Conclusions}
We have presented yields and post-AGB surface abundances for a 1.5\ms\ model of metallicity $Z=0.008$, evolved with three different mass loss laws. We find the mass-loss law has little effect on the mass of the remnant or on the total amount of material dredged-up for this mass of star and metallicity. We also found efficient third dredge-up even at low envelope mass. 

We find a 15\% difference in the carbon yields of the three models and differences of around 20-80\% in the yields of many of the other light elements produced in this mass of AGB star. Post-AGB surface abundances for these models tend to be more uncertain due to dilution effects associated with dredging material into envelopes of decreasing mass.

Comparison of the model abundances to observations is mostly favourable, with many of the model abundances falling with the observed ranges. However, there are some discrepancies. If some additional mixing process were operating, partial CN-cycling at the base of the convective envelope might occur and so could reduce the amount of carbon produced by the models. This could go some way to reconciling the models with the observed C/O ratios, as well as observations of \el{19}{F}.

We stress that this work has only examined one mass of star and one metallicity and our conclusions only apply to these limited circumstances. Future work should expand the comparison over a range of masses and metallicities.

\section{Acknowledgements}
The authors thank the referee, John Lattanzio, for his helpful comments. RJS thanks Armagh Observatory for allowing access to their computing cluster in order to run the simulations presented in this work. He also thanks Churchill College for his fellowship and Ross Church for useful discussions.

\bibliography{masterbibliography}

\appendix
\section{Model details}
Here we present the yields and post-AGB surface abundances (by mass fraction) of all the isotopes in the nucleosynthesis network. The yields are in Table~\ref{tab:yields} and the post-AGB abundances are in Table~\ref{tab:surfaceabundance}.

\begin{table*}
\begin{center}
\begin{tabular}{lcccccccccc}
Mass loss & \el{1}{H} & \el{4}{He} & \el{12}{C} & \el{14}{N} & \el{16}{O} & \el{20}{Ne} & & & \\
\hline
 Reimers & -1.380(-2) & 2.539(-2) & 4.524(-3) & 3.849(-4) & 1.197(-4) & 1.124(-4) & & & &  \\
 Bl\"ocker & -1.063(-2) & 2.676(-2) & 4.449(-3) & 3.911(-4) & 1.155(-4) & 1.148(-4) & & & & \\
VW & -2.264(-2) & 1.885(-2) & 3.884(-3) & 3.748(-4) & 1.234(-4) & 1.098(-4) & \\
\hline
Mass loss & \el{2}{H} & \el{3}{He} & \el{7}{Li} & \el{7}{Be} & \el{11}{B} & \el{13}{C} & \el{14}{C} & \el{15}{N} & \el{17}{O} & \el{18}{O} \\
\hline
 Reimers & -1.445(-5) & 3.718(-4) & -2.786(-9) & 1.121(-15) & -1.012(-9) &  2.241(-5) & 1.133(-7) & -6.306(-7) & 8.512(-6) & -1.441(-6) \\
 Bl\"ocker & -1.475(-5) & 3.799(-4) & -2.844(-9) & 6.019(-15) & -1.033(-9) &  2.285(-5)  & 8.525(-8) & -6.423(-7) & 8.692(-6) & -1.468(-6) \\
 VW & -1.412(-5) & 3.647(-4) & -2.722(-9) & 1.215(-12) & -9.878(-10) & 2.193(-5) &  5.111(-8) & -6.137(-7) & 8.341(-6) & -1.396(-6) \\ 
\hline
Mass loss & \el{19}{F} & \el{21}{Ne} & \el{22}{Ne} & \el{22}{Na} &\el{23}{Na} & \el{24}{Mg} & \el{25}{Mg} & \el{26}{Mg} & \el{26}{Al}$^m$ & \el{26}{Al}$^g$ \\
\hline
Reimers & 3.730(-7) & 3.571(-8) & 2.603(-4) & 0.000 & 6.301(-6) & 1.227(-6) & 2.138(-6) & 1.255(-6) & 0.000 & 6.104(-8) \\
Bl\"ocker & 3.428(-7) & 3.091(-8) & 2.266(-4) & 0.000 & 5.419(-6) & 1.229(-6) &  1.506(-6) & 9.753(-7) & 0.000 &  6.081(-8) \\ 
VW & 2.367(-7) & 1.522(-8) & 1.646(-4) & 4.283(-13) & 4.016(-6) & 1.024(-6) &  9.254(-7) & 6.970(-7) & 0.000 & 5.498(-8) \\  
\hline
Mass loss & \el{27}{Al} & \el{28}{Si} & \el{29}{Si} & \el{30}{Si} & \el{31}{P} & \el{32}{S} & \el{33}{S} & \el{34}{S} & \el{56}{Fe} & \el{57}{Fe} \\
\hline
Reimers & 1.616(-7) & 1.052(-6) & 1.563(-7) & 1.385(-7) & 3.454(-7) & 4.813(-7) &  3.419(-8) & -2.921(-9) & -1.604(-7)  & 5.925(-7) \\
Bl\"ocker & 1.613(-7) & 1.096(-6) & 1.396(-7) & 1.169(-7) & 3.820(-7) & 4.934(-7) &  3.258(-8) & -3.249(-9) & -1.759(-7) & 5.312(-7) \\ 
VW & 1.656(-7) & 1.068(-6) & 1.201(-7) & 1.005(-7) & 2.911(-7) & 5.268(-7) &  2.812(-8) & 5.055(-9) & 3.672(-7) & 4.619(-7)  \\
\hline
Mass loss & \el{58}{Fe} & \el{59}{Fe} & \el{60}{Fe} & \el{59}{Co} & \el{58}{Ni} &
\el{59}{Ni} & \el{60}{Ni} & \el{61}{Ni}  \\
\hline
Reimers & 1.052(-6) & 0.000 & 6.607(-10) & 2.149(-7) & -1.303(-7) & 2.745(-8) & -2.123(-8) &  4.056(-8)  \\
Bl\"ocker & 9.963(-7) & 0.000 & 3.590(-10) & 2.113(-7) & -1.221(-7) & 2.646(-8) & -2.224(-8) & 3.669(-8) & \\
VW & 7.762(-7) & 0.000 & 1.025(-10) & 1.607(-7) & -8.098(-8) & 2.272(-8) & -1.167(-8) & 3.131(-8)  \\
\hline
\end{tabular}
\end{center}
\caption{Net yield from the TP-AGB phase in solar masses for all the isotopes in the nucleosynthesis network. All yields are expressed as $n(m) = n\times10^{m}$.}
\label{tab:yields}
\end{table*}

\begin{table*}
\begin{center}
\begin{tabular}{lcccccccccc}
Mass loss & \el{1}{H} & \el{4}{He} & \el{12}{C} & \el{14}{N} & \el{16}{O} & \el{20}{Ne} & & &   \\
\hline
Reimers & 6.447(-1) &  3.188(-1) & 2.636(-2) & 9.446(-4) & 4.393(-3) & 7.975(-4) & & & & \\
Bl\"ocker & 6.720(-1) & 3.030(-1) & 1.602(-2) & 9.410(-4) & 4.179(-3) & 7.972(-4) & & & \\
VW & 3.182(-1) & 5.021(-1) & 1.579(-1) & 8.081(-4) & 7.867(-3) & 7.998(-4) & & &  \\ \hline
Initial abundance & 7.285(-1) & 2.633(-1) & 1.392(-3) & 4.505(-4) & 3.879(-3) & 7.970(-4) \\
\hline
Mass loss & \el{2}{H} & \el{3}{He} & \el{7}{Li} & \el{7}{Be} & \el{11}{B} & \el{13}{C} & \el{14}{C} & \el{15}{N} & \el{17}{O} & \el{18}{O} \\
\hline
 Reimers & 0.000 & 4.666(-4) &  0.000 & 0.000 & 4.898(-10) & 4.318(-5) & 2.964(-9) & 8.539(-7) & 1.222(-5) & 6.372(-6) \\
 Bl\"ocker & 0.000 & 4.866(-4) & 0.000 & 0.000 &  5.088(-10) & 4.405(-5) & 1.285(-8) & 8.791(-7) & 1.256(-5) & 6.572(-6) \\
 VW & 0.000 &  2.235(-4) & 6.492(-11) & 1.193(-9) & 2.387(-10) & 2.717(-5) &  4.613(-6) & 5.339(-7) & 7.701(-6) & 3.827(-6) \\
\hline
Initial abundance & 1.931(-5) & 1.178(-5) & 3.763(-9) & 0.000 & 1.901(-9) & 1.468(-5) &  0.000 & 1.755(-6) & 1.563(-6) & 8.718(-6) \\
\hline
Mass loss & \el{19}{F} & \el{21}{Ne} & \el{22}{Ne} & \el{22}{Na} &\el{23}{Na} & \el{24}{Mg} & \el{25}{Mg} & \el{26}{Mg} & \el{26}{Al}$^m$ & \el{26}{Al}$^g$ \\
\hline
Reimers & 2.859(-6) & 1.947(-6) & 1.884(-3) & 0.000 & 5.112(-5) & 2.100(-4) &  4.514(-5) & 4.017(-5) & 0.000 & 2.031(-7) \\
Bl\"ocker & 1.418(-6) & 1.777(-6) & 9.047(-4) & 0.000 & 2.955(-5) & 2.082(-4) &  3.384(-5) & 3.471(-5) & 0.000 & 1.314(-7) \\
VW & 1.436(-5) & 3.128(-6) & 9.985(-3) & 4.215(-10) & 1.772(-4) & 2.193(-4) &  1.200(-4) & 7.703(-5) & 0.000 & 9.671(-7) \\
\hline
Initial abundance & 1.629(-7) & 1.660(-6) & 5.238(-5) & 0.000 & 1.343(-5) & 2.071(-4) & 2.722(-5) & 3.122(-5) & 0.000 & 0.000 \\
\hline
Mass loss & \el{27}{Al} & \el{28}{Si} & \el{29}{Si} & \el{30}{Si} & \el{31}{P} & \el{32}{S} & \el{33}{S} & \el{34}{S} & \el{56}{Fe} & \el{57}{Fe} \\
\hline
Reimers & 2.348(-5) & 2.619(-4) & 1.448(-5) & 1.020(-5) & 3.406(-5) & 1.578(-4) &  1.464(-6) & 7.287(-6) & 4.563(-4) & 1.477(-5) \\
Bl\"ocker & 2.346(-5) & 2.623(-4) & 1.410(-5) & 9.785(-6) & 3.356(-5) & 1.584(-4) &  1.392(-6) & 7.383(-6) & 4.625(-4) & 1.326(-5) \\
VW & 2.434(-5) & 2.582(-4) & 1.780(-5) & 1.363(-5) & 4.006(-5) & 1.509(-4) &  2.318(-6) & 6.211(-6) & 3.882(-4) & 3.130(-5) \\
\hline
Initial abundance & 2.332(-5) & 2.627(-4) & 1.378(-5) & 9.463(-6) & 3.281(-5) & 1.592(-4) & 1.296(-6) & 7.507(-6) & 4.703(-4) & 1.148(-5) \\
\hline
Mass loss & \el{58}{Fe} & \el{59}{Fe} & \el{60}{Fe} & \el{59}{Co} & \el{58}{Ni} &
\el{59}{Ni} & \el{60}{Ni} & \el{61}{Ni} &  &  \\
\hline
Reimers & 8.058(-6) &  0.000 & 6.439(-9) & 2.688(-6) & 1.858(-5) & 1.340(-7) &  7.560(-6) & 5.720(-7) &  \\
Bl\"ocker & 4.883(-6) & 0.000 & 1.929(-9) & 2.047(-6) & 1.916(-5) & 8.720(-8) &  7.692(-6) & 4.678(-7) &  \\
VW & 3.919(-5) & 0.000 & 2.647(-8) & 8.813(-6) & 1.211(-5) & 9.557(-7) & 6.026(-6) &  1.718(-6)  \\
\hline
Initial abundance & 1.487(-6) & 0.000 & 0.000 & 1.351(-6) & 1.989(-5) & 0.000 &  7.877(-6) & 3.457(-7) \\
\hline
\end{tabular}
\end{center}
\caption{Final surface abundances (by mass fraction) of all the isotopes in the nucleosynthesis network. All abundances are expressed as $n(m) = n\times10^{m}$. The initial abundances at the time the star formed are included for comparison.}
\label{tab:surfaceabundance}
\end{table*}

\label{lastpage}

\end{document}